\documentclass[floats,floatfix,showpacs,amssymb,prl,twocolumn,superscriptaddress,nofootinbib]{revtex4-1}

\newlength{\figw} 
\usepackage{graphicx,epsf, epsfig, amssymb}
\usepackage{bm}
\usepackage{longtable,tabularx}
\usepackage{color}
\usepackage[breaklinks]{hyperref}
\usepackage{amsfonts,amsmath,amssymb,mathrsfs,gensymb}
\usepackage{natbib}
\usepackage{aas_macros}
\usepackage{array}
\usepackage{multirow}
\usepackage{rotating,array}
\usepackage{graphicx}
\usepackage[normalem]{ulem}

\newcommand\optional[1]{}
\def\be{\begin{equation}}
\def\ee{\end{equation}}
\def\beq{\begin{eqnarray}}
\def\eeq{\end{eqnarray}}
\usepackage{comment}

\begin{document}

\title{Precessional instability in binary black holes with aligned spins}

\author{Davide Gerosa}
\email{d.gerosa@damtp.cam.ac.uk}
\affiliation{Department of Applied Mathematics and Theoretical Physics, Centre for Mathematical Sciences, University of Cambridge, Wilberforce Road, Cambridge CB3 0WA, UK}

\author{Michael Kesden}
\email{kesden@utdallas.edu}
\affiliation{Department of Physics, The University of Texas at Dallas, Richardson, TX 75080, USA }
\author{Richard O'Shaughnessy}
\email{rossma@rit.edu}
\affiliation{Center for Computational Relativity and Gravitation, Rochester Institute of Technology, Rochester, NY 14623, USA}

\author{Antoine Klein}
\email{aklein@olemiss.edu}
\affiliation{Department of Physics and Astronomy, The University of
Mississippi, University, MS 38677, USA}

\author{\\Emanuele Berti}
\email{eberti@olemiss.edu}
\affiliation{Department of Physics and Astronomy, The University of 
Mississippi, University, MS 38677, USA}
\affiliation{CENTRA, Departamento de F\'isica, Instituto Superior
T\'ecnico, Universidade de Lisboa, Avenida Rovisco Pais 1,
1049 Lisboa, Portugal}

\author{Ulrich Sperhake}
\email{u.sperhake@damtp.cam.ac.uk}
\affiliation{Department of Applied Mathematics and Theoretical Physics, Centre for Mathematical Sciences, University of Cambridge, Wilberforce Road, Cambridge CB3 0WA, UK}
\affiliation{Department of Physics and Astronomy, The University of
Mississippi, University, MS 38677, USA}
\affiliation{California Institute of Technology, Pasadena, CA 91125, USA}

\author{Daniele Trifir\`o}
\email{daniele.trifiro@ligo.org}
\affiliation{Department of Physics and Astronomy, The University of 
Mississippi, University, MS 38677, USA}
\affiliation{Dipartimento di Fisica E. Fermi, Universit\`a di Pisa, Pisa 56127, Italy}

\pacs{04.25.dg, 04.70.Bw, 04.30.-w}

\date{\today}

\begin{abstract}
Binary black holes on quasicircular orbits with spins aligned with their orbital angular momentum have been test beds for
analytic and numerical relativity for decades, not least because symmetry ensures that such configurations are
equilibrium solutions to the spin-precession equations.  In this work, we show that these solutions can be unstable when
the spin of the higher-mass black hole is aligned with the orbital angular momentum and the spin of the lower-mass black
hole is antialigned.  Spins in these configurations are unstable to precession to large misalignment when the binary
separation $r$ is between the values $r_{\rm ud\pm}= (\sqrt{\chi_1} \pm \sqrt{q \chi_2})^4 (1-q)^{-2} M$, where $M$ is the
total mass,
$q \equiv m_2/m_1$ is the mass ratio, and $\chi_1$ ($\chi_2$) is the dimensionless spin of the more (less) massive black
hole. This instability exists for a wide range of spin magnitudes and mass ratios and can occur in the
strong-field regime near the merger.  We describe the origin and nature of the instability using recently developed analytical
techniques to characterize fully generic spin precession. This instability provides a channel to circumvent astrophysical spin
alignment at large binary separations, allowing significant spin precession prior to merger affecting both gravitational-wave
and electromagnetic signatures of stellar-mass and supermassive binary black holes.
\end{abstract}
\maketitle 

\noindent{\em Introduction.~--~} Black holes (BHs) have been observed in two distinct regimes: stellar-mass BHs
($5 M_\odot \lesssim m \lesssim 100 M_\odot$) accrete from companions in x-ray binaries \cite{1965Sci...147..394B,
1972Natur.235..271B, 1972Natur.235...37W}, while supermassive BHs shine as quasars or active galactic nuclei (AGN)
\cite{1963MNRAS.125..169H, 1969Natur.223..690L}.  Both types of BHs naturally occur in binaries: the massive stellar
progenitors of stellar-mass BHs are typically formed in binaries, while supermassive BHs form binaries following the mergers
of their host galaxies \cite{1980Natur.287..307B}.  Gravitational radiation circularizes the orbits of these binaries
\cite{1963PhRv..131..435P} and causes them to inspiral and eventually merge, making them promising sources of
gravitational waves (GWs) for current and future GW detectors \cite{2010CQGra..27h4006H, 2013IJMPD..2241010U,
2012CQGra..29l4007S, 2010CQGra..27s4002P, 2013CQGra..30v4010M, 2013PASA...30...17M, 2009arXiv0909.1058J,
2013CQGra..30v4009K}.  The spins of these binary BHs need not be aligned with their orbital angular momentum:
stellar-mass BHs may recoil during asymmetric collapses tilting their spins with respect to the orbital plane
\cite{2000ApJ...541..319K, 2002MNRAS.329..897H, 2013PhRvD..87j4028G}, while the initial orbital plane of supermassive
BH binaries reflects that of their host galaxies and is thus independent of their spin.  Gravitational effects alone will not align
the BH spins with the orbital angular momentum \cite{2004PhRvD..70l4020S, 2007ApJ...661L.147B},
but astrophysical mechanisms exist that drive
the BH spins towards alignment in both regimes.  The first BH to collapse in stellar-mass BH binaries may accrete in a disk
from its as yet uncollapsed companion, while both members of a supermassive BH binary may accrete from a common
circumbinary disk.  Warps in these accretion disks can align the BH spins with the orbital angular momentum
\cite{1975ApJ...195L..65B, 2013ApJ...774...43M,2015MNRAS.451.3941G}, but if the initial misalignment between the BH spin
and accretion disk is greater than $90^\circ$, the BH may instead be driven into antialignment \cite{2005MNRAS.363...49K}.

Misaligned spins cause the orbital angular momentum to precess \cite{1979GReGr..11..149B, 1985PhRvD..31.1815T,
1995PhRvD..52..821K}, modulating the emitted GWs \cite{1994PhRvD..49.6274A}.  
Spin misalignment is both a
blessing and a curse for GW data analysis: it increases the parameter space of templates needed to detect GWs via
matched filtering but also breaks degeneracies between estimated parameters in detected events
\cite{1994PhRvD..49.2658C}.
Misaligned spins at merger can generate large gravitational recoils \cite{2007PhRvL..98w1101G, 2007PhRvL..98w1102C, 2007ApJ...659L...5C}, ejecting supermassive BHs from their host galaxies.  Spin precession may also be responsible for the observed “X-shaped” morphology of AGN radio lobes  \cite{2002Sci...297.1310M,2009ApJ...697.1621G}.
Given the importance of spin misalignment, it is worth investigating the robustness of
aligned spin configurations.  In the general case that the BHs have unequal masses, there are four distinct (anti-)aligned
configurations, which we refer to as up-up, up-down, down-up, and down-down.  The direction before (after) the hyphen
describes the more (less) massive BH and up (down) implies (anti-)alignment of the spin with the orbital angular momentum.
\begin{figure*}[t]
\centering
\includegraphics[width=0.95\textwidth]{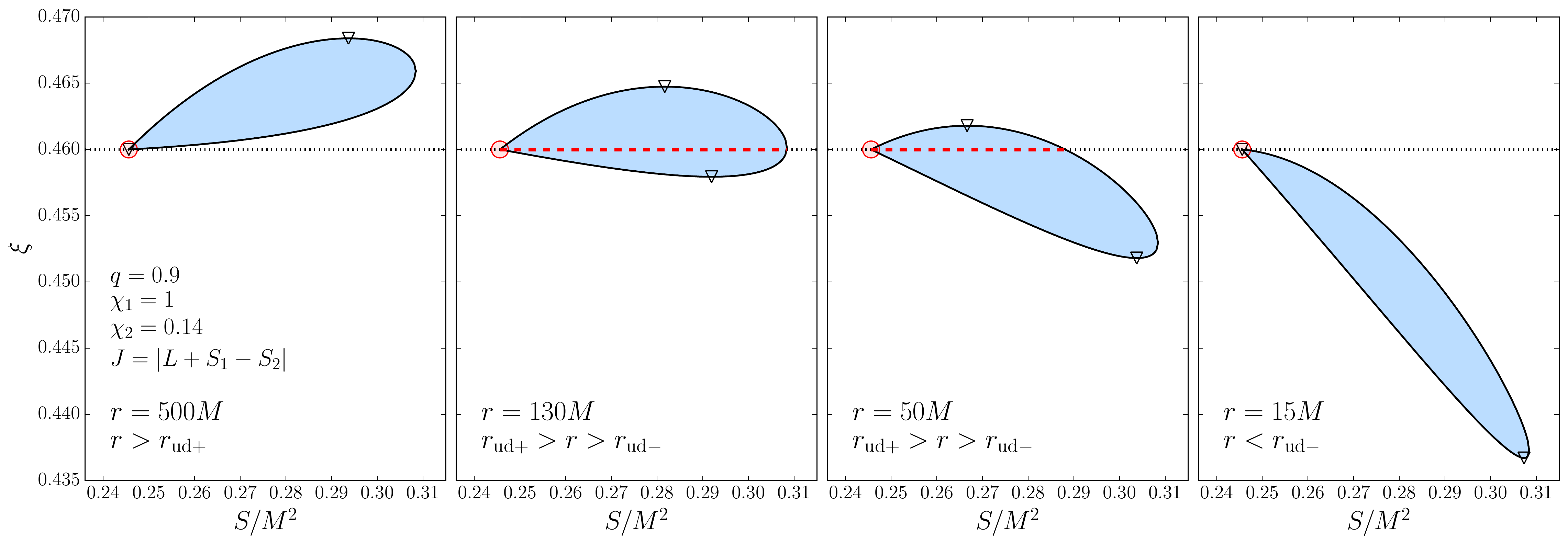}
\caption{Effective-potential loops $\xi_\pm(S)$ for binary BHs with mass ratio $q = 0.9$, dimensionless spins $\chi_1=1$,
$\chi_2=0.14$, and total angular
momentum $J=|L+S_1-S_2|$, corresponding to the up-down configuration.  For binary separations $r > r_{\rm ud+} \simeq
337 M$ (left panel), the up-down configuration at $S_{\rm min}$ marked by a red circle is also a minimum (marked by the
lower triangle).  At intermediate separations $r_{\rm ud+} > r > r_{\rm ud-} \simeq 17 M$ (middle panels), misaligned binaries
with the same value of the conserved $\xi$ exist along the dashed red line.  Perturbations $\delta J$, $\delta\xi$ will cause
$S$ to oscillate between the points $S_\pm$ where this line intersects the loop, making the up-down
configuration unstable.  For $r < r_{\rm ud-}$ (right panel), the up-down configuration is again a stable extremum, now a
maximum (marked by the upper triangle). An animated version of this figure is available online at Ref.~\cite{DGwebsite}.} 
\label{effpot_aligned}
\end{figure*}
By symmetry, all four configurations are
equilibrium solutions to the orbit-averaged spin-precession equations \cite{1995PhRvD..52..821K}, but are these solutions
stable?  To answer this question, we investigate how the configurations respond to perturbations of the spin
directions using our recently developed approach for studying generically precessing systems
\cite{2015PhRvL.114h1103K, 2015arXiv150603492G}.   As we will demonstrate below, the up-down configuration is
unstable for certain choices of binary parameters, with significant consequences for GW data analysis and astrophysics.

\noindent{\em Generic spin precession.~--~} Here we briefly summarize the approach to spin precession
described in detail in \cite{2015PhRvL.114h1103K,2015arXiv150603492G} using units where $G = c = 1$.
Binary BHs with total mass $M = m_1 + m_2$, mass ratio $q =m_2/m_1 \leq 1$, symmetric mass ratio $\eta = q/(1+q)^2$,
and spins $\mathbf{S}_i = \chi_i m_i^2 \hat{\mathbf{S}}_i$
evolve on three distinct time scales: the orbital time $t_{\rm orb} = (r^3/M)^{1/2}$ on which their
separation $\mathbf{r}$ changes direction, the precession time $t_{\rm pre} = (t_{\rm orb}/\eta)(r/M)$ on which the spins and
orbital angular momentum $\mathbf{L}$ change direction, and the radiation-reaction time $t_{\rm RR} =
(t_{\rm orb}/\eta)(r/M)^{5/2}$ on which the magnitudes $r$ and $L$ decrease due to GW emission.  The relative orientations of the spins are often specified by the two angles $\cos\theta_i= {\mathbf{\hat S_i}}
\cdot{\mathbf{\hat L}}$ and the angle $\Delta\Phi$ between the projections of the two spins onto the orbital plane, all  of 
which vary on $t_{\rm pre}$.  The spin orientations can
equivalently be specified by the magnitudes of the total spin $\mathbf{S} 
= \mathbf{S}_1 + \mathbf{S}_2$, the total angular
momentum $\mathbf{J} = \mathbf{L} + \mathbf{S}$, and the projected effective spin \cite{2001PhRvD..64l4013D,
2008PhRvD..78d4021R} $\xi \equiv M^{-2}[(1+q)\mathbf{S}_1 +(1+q^{-1})\mathbf{S}_2] \cdot \mathbf{\hat L}$.
This specification has the advantage that only $S$ evolves on $t_{\rm pre}$, while $J$ evolves on $t_{\rm RR}$ and $\xi$
is conserved throughout the post-Newtonian (PN) stage of the inspiral ($r \gtrsim 10M$) by orbit-averaged 2PN spin
precession and 2.5PN radiation reaction \cite{2010PhRvD..81h4054K}.  
On the precession time, the spin magnitude $S$ simply oscillates back and forth between the two roots $S_\pm$ of the
equation $\xi = \xi_\pm(S)$, where
\begin{align} \label{effpot}
&\xi_\pm(S) = \{ (J^2 - L^2 - S^2)[S^2(1+q)^2 - (S_1^2 - S_2^2)(1- q^2)] \notag \\
& \quad \pm (1- q^2) \sqrt{ [J^2 - (L - S)^2] [(L + S)^2 - J^2]}  \notag \\
& \quad  \times \sqrt{[S^2 - (S_1 - S_2)^2][(S_1 + S_2)^2 - S^2]}\}\big/(4qM^2S^2L)\,,
\end{align}
are the effective potentials for BH spin precession.  Note that $S$ is the only quantity on the right-hand side of
Eq.~(\ref{effpot}) changing on $t_{\rm pre}$; in the absence of radiation reaction, the spins return to their initial relative
orientation after a time $\tau(L, J, \xi)$ during which $\mathbf{L}$, $\mathbf{S}_1$, and $\mathbf{S}_2$ precess about
$\mathbf{J}$ by an angle $\alpha(L, J, \xi)$.
The two potentials $\xi_\pm(S)$ form a closed loop in the $S\xi$ plane, implying that the two roots $S_\pm$ coincide at the
extrema $\xi_{\rm min,max}(L, J)$ of the loop.   
At these extrema, also known as spin-orbit resonances  \cite{2004PhRvD..70l4020S}, $S$ does not oscillate and
$\mathbf{L}$, $\mathbf{S}_1$, and $\mathbf{S}_2$ all remain coplanar on the precession time. 

\noindent{\em Stability of aligned configurations.~--~} 
We begin with the up-up and down-down configurations,
for which $J = |L \pm (S_1+S_2)|$, respectively.  According to Eq.~(\ref{effpot}), the effective-potential loop reduces to a
single point in this limit which is necessarily an extremum: $S$ cannot oscillate consistent with conservation of $J$ and
$\xi$.  Now consider the down-up (up-down) configurations for which $J = |L - S_1 + S_2|$ ($J = |L + S_1 - S_2|$).  The
effective-potential loop $\xi_\pm(S)$ encloses a nonzero area for these values of $J$, implying that oscillations in $S$ are
possible, except at the extrema $\xi_{\rm min,max}$.  Since the spins are antialigned with each other in both configurations,
$S$ is minimized at $S_{\rm min} = |S_1 - S_2|$ and both configurations sit on the leftmost point of the loop, where
$\xi_+(S)$ and $\xi_-(S)$ coincide.  Whether this point is also an extremum $\xi_{\rm min,max}$ depends on the slopes of
these two functions at that point.  Both slopes are always negative for the down-up configuration, implying that it is a
maximum
$\xi_{\rm max}$ and thus a spin-orbit resonance like the up-up and down-down configurations.  At large binary separations
$r$, the slopes of $\xi_\pm(S)$ are both positive for the up-down configuration, making it a minimum $\xi_{\rm min}$.
However, below $r_{\rm ud+}$ given by
\begin{align} \label{E:rud}
r_{\rm ud\pm}= \frac{(\sqrt{\chi_1} \pm \sqrt{q \chi_2})^4}{(1-q)^2}M~,
\end{align} 
the slope of $\xi_-(S)$ becomes negative and up-down is no longer an extremum of the
effective-potential loop, as seen in Fig.~\ref{effpot_aligned}.  At separations below $r_{\rm ud-}$, the slope of $\xi_+(S)$ also
becomes negative and up-down is again an extremum, this time a maximum $\xi_{\rm max}$.  
Misaligned BHs with the {\it same} values of $J$ and $\xi$ as the up-down configuration but $S > S_{\rm min}$ exist in the
intermediate range $r_{\rm ud-} < r < r_{\rm ud+}$, as shown by the dashed red line.  These misaligned BHs have an infinite
precessional period $\tau$: they exponentially approach the up-down configuration on the precession time $t_{\rm pre}$ but
never reach it. 

\begin{figure}[t]
\centering
\includegraphics[height=0.36\textwidth]{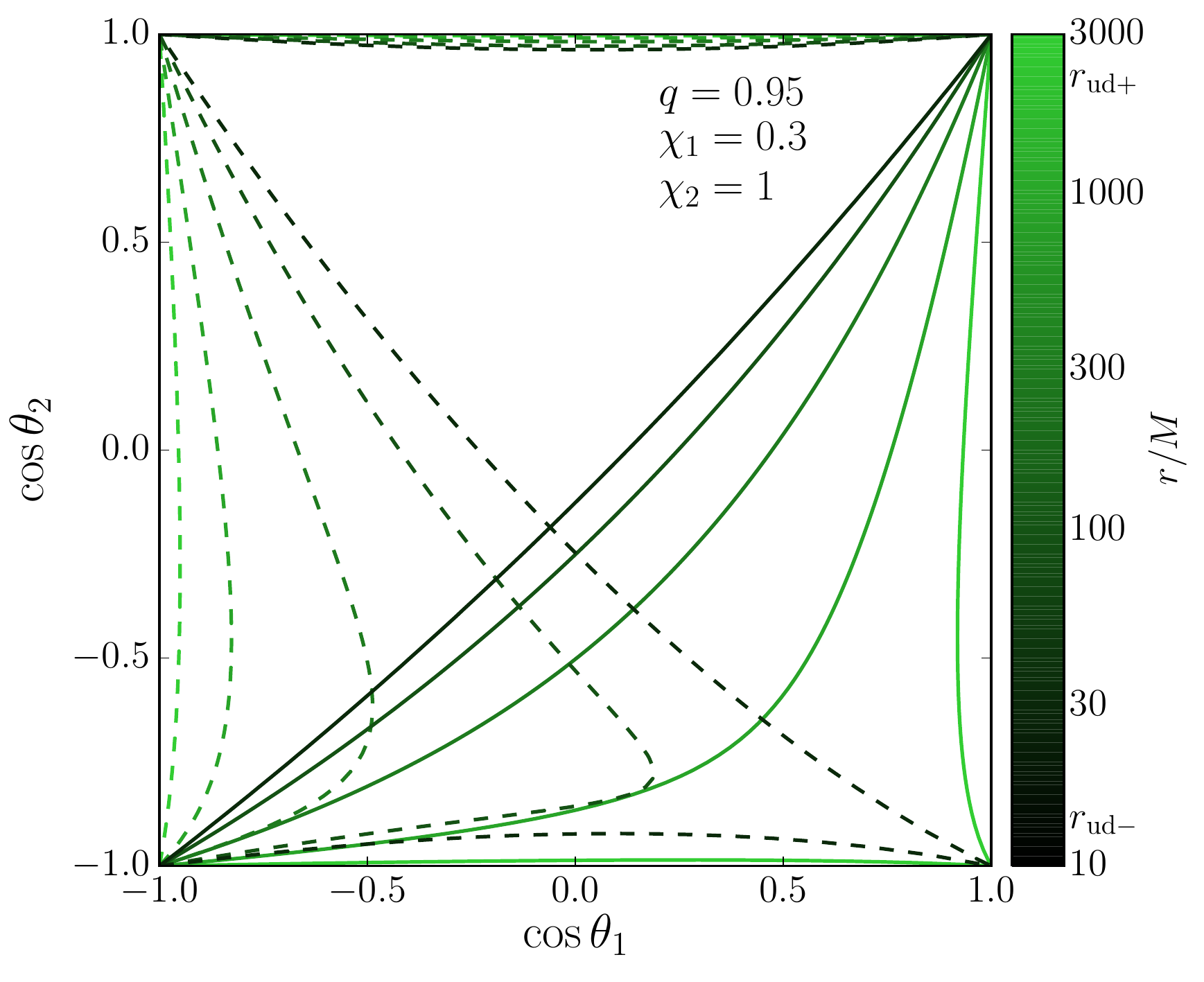}
\caption{The angles $\cos\theta_i= {\mathbf{\hat S_i}} \cdot{\mathbf{\hat L}}$ for spin-orbit resonances [extrema of
$\xi_\pm(S)]$ for BHs with $q = 0.95$, $\chi_1 = 0.3$, and $\chi_2 = 1$.  The solid (dashed) curves indicate the
$\Delta\Phi = 0~(\pi)$ family and the five curves for each family correspond to binary separations $r/M = 3000, 720, 170,
40,\,{\rm and}\,10$.  The up-down configuration (bottom right corner) belongs to the $\Delta\Phi = 0$ family for
$r > r_{\rm ud+} \simeq 2149 M$, to the $\Delta\Phi = \pi$ family for $r < r_{\rm ud-} \simeq 13 M$, and is unstable for
intermediate values $r_{\rm ud-} < r < r_{\rm ud+}$. An animated version of this figure is available online at Ref.~\cite{DGwebsite}.}
\label{rescurves}
\end{figure}

The evolving relationship between the up-down configuration and the spin-orbit resonances parameterized by the angles $\theta_i$ is seen in Fig.~\ref{rescurves}.  The solid curves show the $\Delta\Phi = 0$
resonances [$\xi_{\rm min}(J)$] for separations $10 M \leq r \leq 3000 M$, while the dashed curves show the $\Delta\Phi =
\pi$ resonances [$\xi_{\rm max}(J)$].  The up-down configuration is located in the bottom right corner of this figure. 
For $r > r_{\rm ud+}$, the up-down configuration lies on the solid curves and belongs to the $\Delta\Phi = 0$ family, but for
smaller separations these curves detach from the bottom right corner, and thus up-down is no longer a minimum of
$\xi_\pm(S)$.  The dashed curves indicating the $\Delta\Phi = \pi$ family migrate to the right with decreasing separation and
reach the bottom right corner, making the up-down configuration a
maximum of $\xi_\pm(S)$, for $r < r_{\rm ud-}$.  The up-up and down-down configurations (top right and bottom left corners) belong to both resonant families,
reflecting the degeneracy of the effective-potential loop as a single point that is both minimum and maximum.  The down-up
configuration (top left) always belongs to the $\Delta\Phi = \pi$ family and is thus a maximum $\xi_{\rm max}$.

\begin{figure}[t]
\centering
\includegraphics[height=0.36\textwidth]{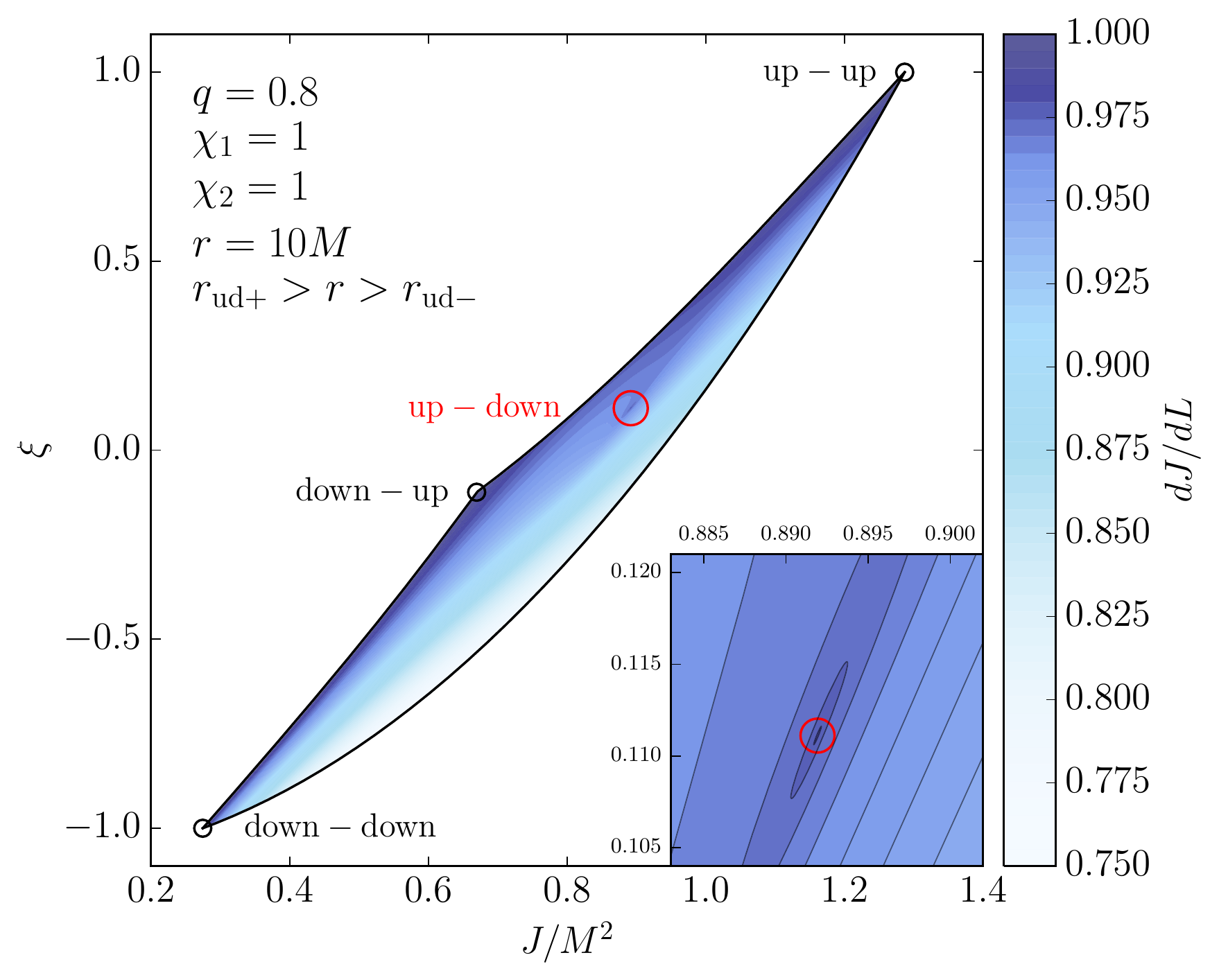}
\caption{Precession-averaged radiation reaction $dJ/dL$ as a function of $J$ and $\xi$ for binaries with $q = 0.8$, $\chi_1
= \chi_2 = 1$, and separation $r = 10M$ in the unstable region $r_{\rm ud-} < r < r_{\rm ud+}$.  Spin-orbit resonances
including the up-up, down-down, and down-up configurations are extrema of $\xi_\pm(S)$ and constitute the boundary of
the allowed region.  All four aligned configurations are maxima where $dJ/dL = 1$, but the unstable up-down configuration
(shown in the inset) is a cusp. 
An animated version of this figure is available online at Ref.~\cite{DGwebsite}.}
\label{dJdLcont}
\end{figure}

The stability of a system is determined by its response to perturbations, in this case to the spin angles ($\delta\theta_1,
\delta\theta_2, \delta\Delta\Phi$) or equivalently to the angular momenta ($\delta S, \delta J, \delta\xi$).  After such a
perturbation, configurations that are extrema of $\xi_\pm(S)$ (all aligned configurations except up-down for $r_{\rm ud-} < r
< r_{\rm ud+}$) will undergo oscillations in $S$ (and thus the three spin angles) that are linear in the perturbation amplitude,
and have a period $\tau$ that is independent of this amplitude.  This is a stable response equivalent to that of a simple
harmonic oscillator.  The response of the up-down configuration for $r_{\rm ud-} < r < r_{\rm ud+}$ is very different, as seen
in the middle panels of Fig.~\ref{effpot_aligned}: $S$ oscillates between the turning points $S_\pm$ independent of the
perturbation amplitude, but the period $\tau$ of these oscillations --
as predicted by Eq.~(27) of Ref.~\cite{2015arXiv150603492G} --
diverges logarithmically as this amplitude approaches zero.
This is an unstable response: the time it takes for a zero-energy particle with $dx/dt < 0$ to travel from finite $x_0$ to
$\delta x$ in the unstable potential $V = -\frac{1}{2}kx^2$ similarly diverges logarithmically with $\delta x$.

A perturbative analysis of nearly aligned configurations \cite{2013PhRvD..88l4015K} can identify that perturbations can
oscillate at complex frequencies (indicating an instability) in the same region $r_{\rm ud-} < r < r_{\rm ud+}$ found here,
but such analysis cannot predict the amplitude of these perturbations or their response to precession-averaged radiation
reaction.

\noindent{\em Radiation reaction.~--~} We have shown that for $r_{\rm ud-} < r < r_{\rm ud+}$, spin configurations with
$J$ and $\xi$ infinitesimally close to the up-down configuration can experience finite-amplitude oscillations in $S$ and the
angles $\theta_1$, $\theta_2$, and $\Delta\Phi$.  We now investigate how these configurations evolve on the
longer radiation-reaction time $t_{\rm RR}$.  Since $\xi$ is conserved throughout the inspiral and $L$ monotonically
decreases at 2.5PN order, the only challenge is to evolve $J$.  In Refs.~\cite{2015PhRvL.114h1103K,2015arXiv150603492G} we derived a precession-averaged expression for $dJ/dL$, a contour plot of which is shown in Fig.~\ref{dJdLcont}.  The shaded
region shows the allowed values of $J$ and $\xi$ for this mass ratio, spin magnitudes, and binary separation.  The
spin-orbit resonances, being extrema of $\xi_\pm(S)$, constitute the boundaries of this region.  The up-up, down-down, and
down-up configurations, being spin-orbit resonances, lie on these boundaries.  At $r_{\rm ud+}$, the up-down configuration
detaches from the right boundary of this region [it stops being a minimum of $\xi_\pm(S)$] and begins to migrate leftwards
through the allowed region, eventually reattaching to the left boundary at $r_{\rm ud-}$ [where it becomes a maximum of
$\xi_\pm(S)$].  This is just an alternative visualization of the four panels of Fig.~\ref{effpot_aligned}.

\begin{figure}[t]
\centering
\includegraphics[width=0.40\textwidth]{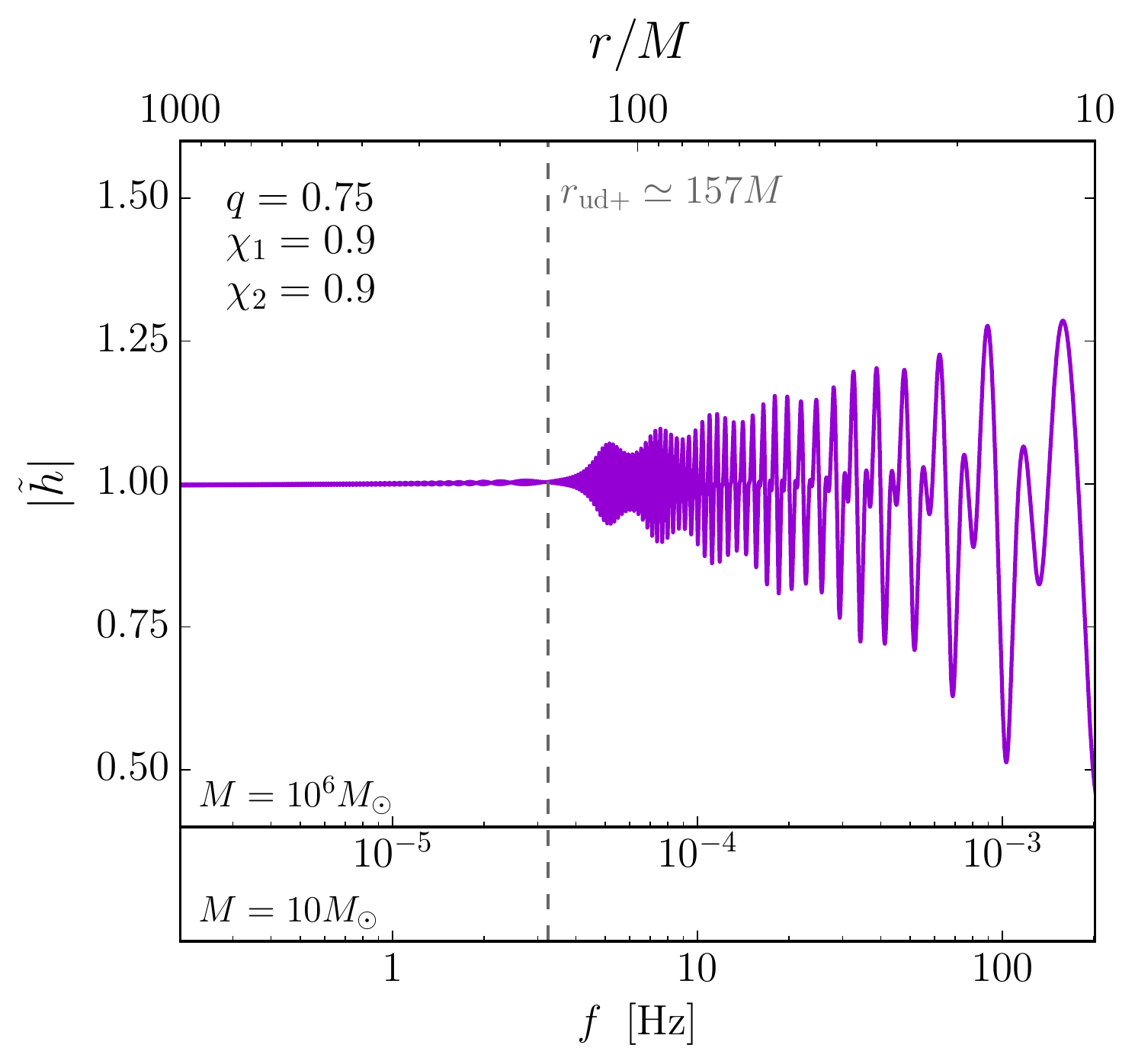}
\caption{Normalized GW Fourier amplitude $\tilde{h}$ (cf. Ref.~\cite{2013PhRvD..88l4015K}) as a function of orbital 
frequency $f$ and binary separation $r$ during the inspiral
of BHs with $q = 0.75$ and $\chi_1 = \chi_2 = 0.9$.
At the initial separation $r = 1000M$, the spins are nearly in the up-down configuration, but this configuration becomes
unstable below $r_{ud+} \simeq 157 M$, after which large precession-induced modulations occur at
frequencies accessible to GW detectors.\vspace{-0.3cm}}
\label{GWfig}
\end{figure}

For all four aligned configurations, $\mathbf{J}$ and $\mathbf{L}$ are aligned so $dJ/dL = 1$ is maximized.  However, the
nature of these maxima is very different for the stable and unstable configurations.   For the stable configurations, the partial
derivatives of $dJ/dL$ with respect to $J$ and $\xi$ remain finite, implying that neighboring points separated by
($\delta J, \delta\xi$) slowly drift away at a rate that scales linearly with these infinitesimal quantities.  The unstable 
configuration however is a cusp where these partial derivatives approach $\pm\infty$, depending on whether this point in
the $J\xi$ plane is approached from
below or above.  Neighboring points (experiencing large-amplitude oscillations in $S$, as seen in the middle panels of
Fig.~\ref{effpot_aligned}) rapidly deviate from the up-down configuration as it sweeps across the allowed region.  This is an
essential point: even if the stability of the up-down configuration is restored in the PN regime ($r_{\rm ud-} > 10 M$),
radiation reaction during the inspiral between $r_{\rm ud\pm}$ will drive BHs initially in this configuration to large
misalignments prior to merger.  The migration of the up-down configuration through the $J\xi$ plane also reconciles the
instability with the empirical result that isotropic spin distributions remain isotropic during the inspiral 
\cite{2004PhRvD..70l4020S, 2007ApJ...661L.147B}: although nearby binaries may indeed be left behind, the unstable
configuration will always encounter a fresh supply, until it is restored to stability at the left edge of the allowed region.

\noindent{\em GW astronomy.~--~} 
Binaries with separations in the unstable region between $r_{\rm ud\pm}$ emit GWs with frequencies in the range
$f_{\rm ud \pm} \simeq 6.4\times 10^4  {\rm Hz} (M/M_\odot)^{-1} (1 - q)^3/(\sqrt{\chi_1}\pm \sqrt{q \chi_2})^{6}$, within or
below the sensitivity band of existing and planned GW detectors \cite{2010CQGra..27h4006H, 2013IJMPD..2241010U,
2012CQGra..29l4007S, 2010CQGra..27s4002P, 2013CQGra..30v4010M, 2013PASA...30...17M, 2009arXiv0909.1058J,
2013CQGra..30v4009K}.  In Fig.~\ref{GWfig}, we show the waveform of one such binary initially near the up-down
configuration before entering the unstable region.  Once the binary crosses the threshold at $r_{\rm ud+}$, its waveform
develops large-amplitude precessional modulation on the precession time $t_{\rm pre}$.  The amplitude of this modulation
is independent of the initial deviation from the up-down configuration: it is set by the finite-amplitude oscillations in $S$ seen
in the middle panels of Fig.~\ref{effpot_aligned}.  
Modulation occurs on two distinct time scales associated with the
precession of $\mathbf{L}$ in a frame aligned with $\mathbf{J}$. In this frame
the direction of $\mathbf{L}$ is specified by
the polar angle $\cos\theta_L = \hat{\mathbf{L}} \cdot \hat{\mathbf{J}}$ and the azimuthal angle $\Phi_L$ in the plane
perpendicular to $\mathbf{J}$.  The longer of these time scales is $\tau$ (the period of oscillations in $\theta_L$), while the
shorter time scale is $(2\pi/\alpha)\tau$ (the precession-averaged time for $\Phi_L$ to change by $2\pi$)
\cite{2015PhRvL.114h1103K, 2015arXiv150603492G}.
Measuring this modulation could yield insights into the astrophysical origins of binary BHs
\cite{2013PhRvD..87j4028G, 2015arXiv150603492G}.  Spin precession could also affect the
electromagnetic counterparts to BH mergers \cite{2005ApJ...622L..93M, 2007PhRvL..99d1103L}
and the probability of ejecting a supermassive BH from its host galaxy \cite{2007PhRvL..98w1101G, 2007PhRvL..98w1102C, 2007ApJ...659L...5C,2010ApJ...715.1006K}.
We look forward to confronting these predictions with observations in the dawning age of GW astronomy.

\noindent{\em Acknowledgments.~--~} We thank Tyson Littenberg for
discussions. D.G. is supported by the UK STFC and the Isaac Newton
Studentship of the University of Cambridge.  M.K.~is supported by
Alfred P. Sloan Foundation Grant No. FG-2015-65299.  R.O'S. is supported
by NSF Grants No. PHY-0970074 and No. PHY-1307429.  A.K. and E.B.~are
supported by NSF CAREER Grant PHY-1055103. 
E.B. acknowledges support from FCT Contract No. IF/00797/2014/CP1214/CT0012 under the IF2014 Programme.
U.S.~is supported by FP7-PEOPLE-2011-CIG Grant No.~293412,
FP7-PEOPLE-2011-IRSES Grant No. 295189, H2020-MSCA-RISE-2015 Grant No.~StronGrHEP-690904, SDSC and TACC through XSEDE
Grant No.~PHY-090003 by the NSF,
H2020 ERC Consolidator Grant Agreement No.~MaGRaTh-646597,
STFC Roller Grant No. ST/L000636/1 and DiRAC's
Cosmos Shared Memory system through BIS Grant No.~ST/J005673/1 and
STFC Grant Nos.~ST/H008586/1, ST/K00333X/1.  
D.T. is partially supported by the NSF awards PHY-1067985 and PHY-1404139.
Figures were generated
using the \textsc{python}-based \textsc{matplotlib} package
\citep{2007CSE.....9...90H}.

\bibliography{updown}
\end{document}